\UseRawInputEncoding

\documentclass{ws-ijmpd}
\usepackage[super,compress]{cite}
\begin{document}


%
%

\title{Neutron stars as probes of dark matter}

\author{  M. \'Angeles P\'erez-Garc\'ia$^1$}
\address{$^1$Department of Fundamental Physics, University of Salamanca, \\ Plaza de la Merced s/n 37008 Spain
}

\author{   Joseph Silk$^{1,2,3}$}
\address{$^2$Institut d'Astrophysique,  UMR 7095 CNRS, Sorbonne Universit\'e, \\98bis Blvd Arago, 75014 Paris, France\\ $^3$Department of Physics and Astronomy, The Johns Hopkins University,\\ Homewood Campus, Baltimore MD 21218, USA\\
$^4$Beecroft Institute of Particle Astrophysics and Cosmology, Department of Physics, \\University of Oxford, Oxford OX1 3RH, UK\\
}

\maketitle

\begin{abstract}

Neutron Stars (NSs) are compact stellar objects that are stable solutions in General Relativity. Their  internal structure is usually described using an equation of state that involves the presence of ordinary matter and its interactions. However there is now a large consensus that an elusive sector of matter in the Universe, described as dark matter, remains as yet undiscovered. In such a case, NSs should contain both, baryonic  and dark matter. We argue that depending on the nature of the dark matter and in certain circumstances, the two matter components would form a mixture inside NSs that could trigger further changes, some of them observable. The very existence of NSs constrains the nature and interactions of dark matter in the Universe.
\\

This essay received an Honorable Mention in the 2020 Essay Competition of the Gravity Research Foundation.
\end{abstract}

\keywords{dark matter, baryonic matter, neutron star.}

\ccode{PACS numbers: 95.35.+d, 21.65.-f, 26.60.-c, 97.60.Jd, 21.30.Fe}


\section{Introduction}

The concept of matter is rooted in our sensorial experience. From its origins in the natural philosophy of Democritus and Aristotle  to the modern perspectives of Newton, Planck and Einstein, matter (energy) characterizes the observable quantities of interest for any physical system. In the case of gravity, matter distorts the otherwise flat space-time creating a curved region described by the theory of General Relativity (GR). By solving Einstein's field equations under various  simplifying assumptions, one can obtain a set of finite-radius stable solutions that can be identified with stellar bodies. Either static or rotating stellar solutions are characterized by a mass ($M$) and radius ($R$). The actual stellar structure depends on the behaviour of matter, i.e. interactions among its constituents. The equation of state (EoS) relates the pressure and energy density at a given temperature and is a crucial ingredient that incorporates  all of this information. Relying on the Standard Model of Particle Physics, the building blocks of ordinary matter involve quarks and leptons, and the interactions are mediated by bosonic force carriers.

Depending on the conditions of the stellar interior i.e. density and temperature, the particular state adopted by matter can be inferred according to our knowledge of the general phase diagram of matter. In this way, low density sun-like stars possess a plasma of light nuclei and electrons in their core  with density $\rho\sim 10^2$ $\rm g/cm^3$. At much higher densities, beyond that of nuclear saturation $\rho\sim 10^{14}$ $\rm g/cm^3$ nucleons ($N$), leptons and  other hadrons yield EoS which produce compact stars, Neutron Stars (NSs), when solving the stellar structure equations \cite{eos}. Recent NS measurements aim to constrain the yet uncertain high density, low temperature matter phase diagram using several techniques including analysis of the gravitational wave pattern emitted in merger events, X-ray emission or pulse profile modeling of the Shapiro delay to determine their masses and radii,  see  Ref. \refcite{nsobservations}.  

However, the aforementioned scenario does not consider the so-called dark matter (DM). This is an elusive ingredient that hitherto remains undetected and is thought to constitute more than $85\%$ of all gravitating matter\cite{reviewdm}. Cosmological-scale phenomena can be described with different approaches and at different degrees of accuracy including DM although some alternatives exist in the literature as well.  For example, that of the cosmic microwave background (CMB) can be accounted for with the beforementioned dark sector but without modifying GR \cite{planck}. Alternative approaches that modify GR are also under current scrutiny since they could also be viable and it that way the CMB, Big Bang nucleosynthesis, and expansion rate history can be accounted for \cite{angus} with the so-called hot DM. Milgromian dynamics (MOND)  were  originally developed to explain galaxy rotation curves, and have had much predictive success in this regard \cite{Famaey}\cite{mcgaugh} .  Further, it is also claimed that it is possible that the Universe does not
contain any DM at all, which would produce an unusually strong redshifted 21 cm absorption feature from the epoch of reionisation \cite{ion}. 

In this essay we will argue that under the prescribed scenario with a dark particle content the current picture of the interior of NSs is far from  complete unless we account for this (cold) DM. When this is done, NSs are  revealed to contain an even richer complexity.  Some of their features, in turn, can constrain some of the dark sector candidates that  already exist in the literature.  It is important to remark that when solving for the structure of NSs, one must account for {\it all} kinds of matter and fields that are involved. Dynamical effects in the gravitational wave pattern that can emerge in merger events such as those of NS-NS or NS-black hole will also serve to constrain the properties of the possible mixture of dark and ordinary (baryonic) matter \cite{gw} or other approaches using modified gravity. It is worth recalling that for isolated NS or those in binaries, DM could also yield substantial and observable differences with respect to DM-free objects.

\section{Dark Matter candidates}

Currently there is a vast variety of DM candidates, for example  weakly or strongly interacting massive particles, sterile neutrinos, axions, and primordial black holes (PBHs), just to mention a few of the most popular \cite{dmcandidates}. We propose here that NSs can shed light on this under the premise that they simply exist. Displaying a core with mostly nucleon content at low temperature, NSs can virtually test mass, cross-sections and the spin-statistical nature of DM in a possible mixture inside the compact star. For this argument, we will build on previous research originally published in \cite{prl} and in a subsequent series of works \cite{daigne, cermeno}.

NSs have typical masses $M\gtrsim2M_\odot$ and radii $R\approx 11-14$ km. They can effectively be opaque to DM particles in a galactic distribution that accretes a small DM fraction. This depends on the kinematics of the incoming DM particle with mass $m_\chi/c^2$ and scattering cross-section with nucleons $\sigma_{\chi N}$. This two-parameter space mostly determines the ability of the star to externally incorporate DM. In this phenomenological exploration, the current experimentally excluded region dramatically increases as the detection limits in the few-GeV$\rm /c^2$ mass range are lowered, and gets close to the coherent neutrino scattering limit. In addition, collider searches provide competitive limits for light DM while for larger masses, hard photons and neutrinos could probe models where heavy DM is produced. As these various options have all been systematically explored with null results,  light DM with $m_\chi\lesssim 1$ $\rm GeV/c^2$ remains a promising candidate. Axions and a more relaxed version of them, the axion-like particles (ALPs),  also yield interesting candidates. Axion searches are  being conducted by terrestrial experiments and constrained by magnetic fields in NSs such as those in magnetars  with $B\lesssim 10^{15}$ G \cite{magnet}. 

\section{DM Trojan Horse mechanism inside NSs}

In this hunt for a  DM smoking gun signal, a variety of new strategies have been developed when considering the NS laboratory. Since they have a large compactness $M/R\sim 0.1$,  NSs are very efficient sinks of DM and have large associated capture rates.  The impact of considering either GR vs Newtonian gravity increases these rates \cite{gould} to around $\sim 70\%$ being 

\begin{equation}
C_{\chi} \simeq 3 \times 10^{28} \left(\frac{M}{1.5 \,M_{\odot}}\right) \left(\frac{R}{10\,\rm km}\right) \left(\frac{1\, \rm MeV/c^2}{m_{\chi}}\right)\left(\frac{\rho^{ambient}_{\chi}}{0.4 \,\rm \frac{GeV}{cm^3}}\right) \left( \frac{\sigma_{\chi N}}{\sigma_0}\right)\,\,\rm s^{-1}.
\end{equation}

  As usual, a reference value for our solar system is $\rho^{ambient}_{\chi,0}\simeq 0.4$ $\rm \frac{GeV}{cm^3}$. The ratio $\frac{\sigma_{\chi N}}{\sigma_0}$ denotes how a tiny cross-section $\chi N$  for a particular model can limit the opacity of NS matter. The geometrical cross-section is defined as $\sigma_0=\frac{m_N R^2}{M}=6 \times 10^{-46}$ $\rm cm^2$ so that this rate saturates for  $\sigma_{\chi N}> \sigma_0$.

For most of the  strongly or weakly interacting DM particles, the number fraction inside an old NS, with a baryon number of $N_B\sim 10^{58}$,  is tiny:   $Y_\chi \equiv N_{\chi}/N_B\lesssim 10^{-16}$ and gravitational distortions are not expected. Nevertheless other mechanisms based on a {\it Trojan Horse} strategy may indeed provide substantial internal changes as we will explain.

For large enough cross-sections,  a thermalized non-relativistic distribution is built inside the core over time \cite{bridget}. The number of DM particles inside the star at a given time is the result of a series of internal processes that may produce \cite{decayX} or remove further particles beyond those that are gravitationally captured and do not evaporate. Majorana particles can self-annihilate or co-annihilate with other species and partially eliminate the reservoir of stellar DM. These candidates inject energy into the cold NS core and are possible catalyzers of further internal thermodynamic  phase transitions. For example,  quark content could be deconfined if the energy to overcome the confining hadronic potential is released \cite{herrero}. How strong a case there is for self-annihilation is unknown, but a 
thermally-averaged value of this cross-section based on {\it freeze-out} arguments at an early epoch of Universe  is usually taken as  $\langle \sigma_a v\rangle\simeq 10^{-26}\, \rm  cm^3\, s^{-1}$.

\section{Observational hints}

 In general, there are several indirect signals that could be linked to DM existence in  NS scenarios. One such observational consequence would be the generation of a very short Gamma Ray Burst (GRB) with a hard spike spectrum and few ms duration \cite{daigne} when the NS transition to a more compact quark star takes place. The  asymmetric DM \cite{zurek, kouvaris}, described as Dirac fermions, will follow exclusion rules that imply the existence of a  Chandrasekhar dark mass, that scales as  $\sim M^3_P$  where $M_P = 1.22 \times 10^{19}$ $\rm GeV/c^2$ is the Planck mass. Numerically,

\begin{equation}
M_{\chi,f} \sim 1.8 \times 10^{57} \left(\frac{1 \,\rm GeV/c^2}{m_\chi}\right)^2.
\end{equation}

This would create a new  possibility that an induced collapse to a BH  takes place due to the DM core reaching the critical mass. This would have a strong impact in the stellar evolution and emission patterns of gravitational waves, and  could be detected as signaling a sharp transition  when the BH forms. Bosonic DM is another possibility for forming a condensate inside the star with a critical mass \cite{ruffini}

\begin{equation}
M_{\chi,b} \sim 9.4 \times 10^{37} \left(\frac{1 \,\rm GeV/c^2}{m_\chi}\right).
\end{equation}

The internal dynamics of the early phases of the proto-NS can also be changed from a standard scenario by certain types of light DM. The interior of the star reaches a temperature $k_BT\sim 50$ MeV immediately after the supernova collapse. The presence of an additional component of light DM with a weak cross-section could facilitate  heat and energy conduction from  the central part of the star to the outer regions. This effect could be observable at later times as a plateau of hotter stars than expected for the NS ages, since heat is conducted more efficiently due to the extra steady flux of incoming, accreting  DM. Possible candidates for such objects are the luminous and apparently young stars detected within 0.1 pc of the galactic centre \cite{ciurlo}, where the DM density is likely to be elevated by the presence of the central supermassive BH \cite{gondolo}.

To summarize, we have discussed an interesting scenario under the current cosmological paradigm where Neutron Stars could be considered as hybrid objects made from standard and cold dark matter. Depending on the nature of the prescribed dark matter particle and its interactions inside the star a number of critical circumstances could lead to the disruption of its regular structure and further observational consequences, thus constituting a probe of the dark sector.

\section*{Acknowledgments}
M. A. P. G. would like to thank partial financial support from University of Salamanca, Junta de Castilla y Le\'on SA083P17, SA096P20, Ministry of Science  PID2019-107778GB-100,  the MULTIDARK Spanish network  and PHAROS COST Action CA16214.




\begin{thebibliography}{0}    

\bibitem{eos} J. M. Lattimer and M. Prakash, Astrophys. J. 550 (2001) 426.
\bibitem{nsobservations} F. Özel and P. Freire,  Annual Review of Astronomy and Astrophysics 54 (2016) 401.
\bibitem{planck} Planck Collaboration, Y. Akrami, F. Arroja, M. Ashdown et al., accepted in Astronomy and Astrophysics (2020) 
\bibitem{angus} G. W. Angus, Monthly Notices of the Royal Astronomical Society, 394 (2009) 527
\bibitem{Famaey}B. Famaey and  S. McGaugh , Living Reviews in Relativity, 15 (2012)10.
\bibitem{mcgaugh} S. McGaugh, F. Lelli and  J. Schombert, Phys. Rev. Lett. 117 (2016) 201101.
\bibitem{ion} S. McGaugh, Phys. Rev. Lett. 121 (2018) 081305.
\bibitem{reviewdm} G. Bertone, D. Hooper, J. Silk, Phys .Rept. 405 (2005) 279.
\bibitem{gw}  B. P. Abbott et al. Phys. Rev. Lett. 121 (2018) 161101.
\bibitem{dmcandidates} J. L. Feng, Ann. Rev. Astron. Astrophys. 48 (2019) 495.
\bibitem{prl} M. A.  P\'erez-Garc\'ia , J. Silk, J. R. Stone, Phys. Rev. Lett. 105 (2010) 141101.
\bibitem{daigne}  M. A.  P\'erez-Garc\'ia, Daigne, F., Silk, J.  2013,  ApJ 768 (2013) 145.
\bibitem{cermeno} M. Cerme\~no, M. A.  P\'erez-Garc\'ia  and J. Silk,  Publications of the Astronomical Society of Australia, 34 (2017) 
\bibitem{magnet} J. Fortin,  K. Sinha,  J. High Energ. Phys. 2018 (2018) 48.
\bibitem {gould} A. Gould,  ApJ 321 (1987) 571.
\bibitem {bridget} B. Bertoni, A. E. Nelsonand S. Reddy, Phys. Rev. D 88 (2013) 123505.
\bibitem {decayX} B. Fornal and B. Grinstein, J. Phys.: Conf. Ser. 1308 (2019) 012010.
\bibitem {herrero} A. Herrero, M. A. Pérez-García, J. Silk, and C. Albertus, Phys. Rev. D 100 (2019) 103019 .
\bibitem{zurek} K. M. Zurek,  Phys. Rept. 537 (2014) 91.
\bibitem{kouvaris} C. Kouvaris, Physical Review D 83 (2010) 083512.
\bibitem{ruffini} R. Ruffini and S. Bonazzola, Phys. Rev. 187 (1969)  1767.
\bibitem{ciurlo} A. Ciurlo et al., Nature 577 (2020) 337.
\bibitem {gondolo} P. Gondolo, J. Silk, Phys. Rev. Lett. 83 (1999) 1719.


\end{thebibliography}
\end{document}